\documentclass[twoside,twocolumn,9pt]{article}
\usepackage{extsizes}
\usepackage[super,sort&compress,comma]{natbib} 
\usepackage[version=3]{mhchem}
\usepackage[left=1.5cm, right=1.5cm, top=1.785cm, bottom=2.0cm]{geometry}
\usepackage{balance}
\usepackage{widetext}
\usepackage{times,mathptmx}
\usepackage{sectsty}
\usepackage{graphicx} 
\usepackage{lastpage}
\usepackage[format=plain,justification=raggedright,singlelinecheck=false,font={stretch=1.125,small,sf},labelfont=bf,labelsep=space]{caption}
\usepackage{float}
\usepackage{fancyhdr}
\usepackage{fnpos}
\usepackage[english]{babel}
\usepackage{array}
\usepackage{droidsans}
\usepackage{charter}
\usepackage[T1]{fontenc}
\usepackage[usenames,dvipsnames]{xcolor}
\usepackage{setspace}
\usepackage[compact]{titlesec}

\usepackage{epstopdf}

\definecolor{cream}{RGB}{222,217,201}

\usepackage{xfrac}

\usepackage{amssymb}
\newcommand{\debye}{\lambda_\text{D}}
\newcommand{\kt}{k_\text{B} T}
\newcommand{\zs}{z_\text{s}}
\newcommand{\zsp}{z_\text{s}'}
\newcommand{\zes}{z_\text{ES}}
\newcommand{\new}[1]{#1}

\begin{document}

\pagestyle{fancy}
\thispagestyle{plain}
\fancypagestyle{plain}{

\fancyhead[C]{\includegraphics[width=18.5cm]{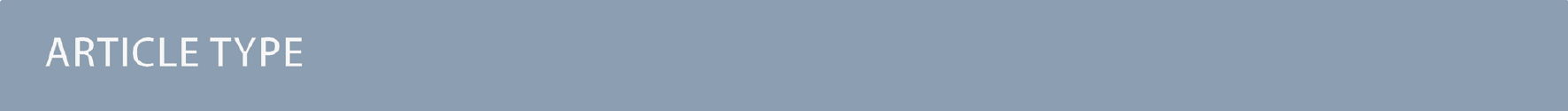}}
\fancyhead[L]{\hspace{0cm}\vspace{1.5cm}\includegraphics[height=30pt]{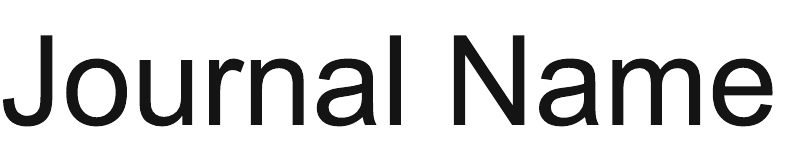}}
\fancyhead[R]{\hspace{0cm}\vspace{1.7cm}\includegraphics[height=55pt]{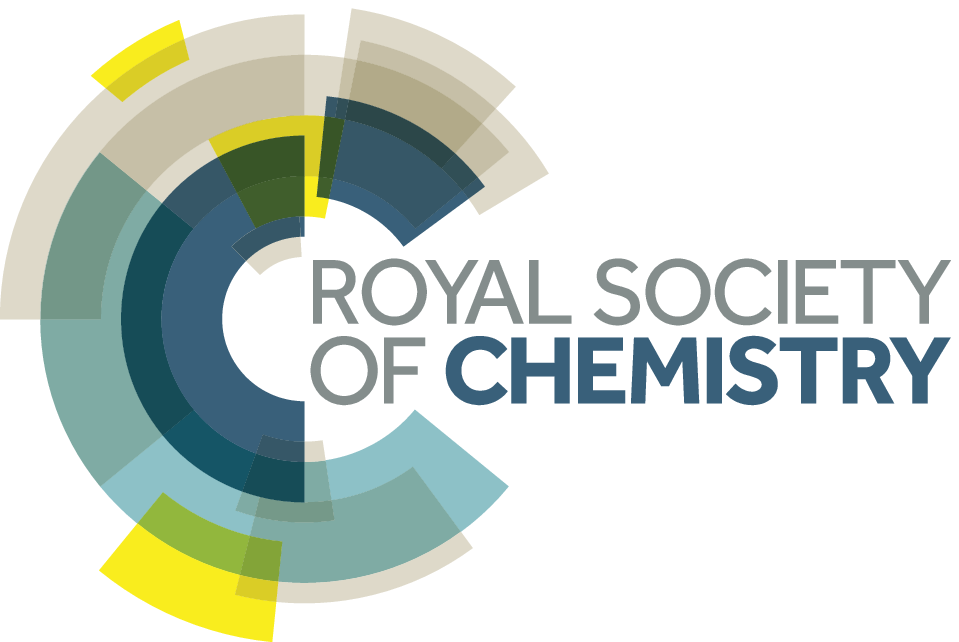}}
\renewcommand{\headrulewidth}{0pt}
}

\makeFNbottom
\makeatletter
\renewcommand\LARGE{\@setfontsize\LARGE{15pt}{17}}
\renewcommand\Large{\@setfontsize\Large{12pt}{14}}
\renewcommand\large{\@setfontsize\large{10pt}{12}}
\renewcommand\footnotesize{\@setfontsize\footnotesize{7pt}{10}}
\makeatother

\renewcommand{\thefootnote}{\fnsymbol{footnote}}
\renewcommand\footnoterule{\vspace*{1pt}%
\color{cream}\hrule width 3.5in height 0.4pt \color{black}\vspace*{5pt}} 
\setcounter{secnumdepth}{5}

\makeatletter 
\renewcommand\@biblabel[1]{#1}            
\renewcommand\@makefntext[1]%
{\noindent\makebox[0pt][r]{\@thefnmark\,}#1}
\makeatother 
\renewcommand{\figurename}{\small{Fig.}~}
\sectionfont{\sffamily\Large}
\subsectionfont{\normalsize}
\subsubsectionfont{\bf}
\setstretch{1.125} 
\setlength{\skip\footins}{0.8cm}
\setlength{\footnotesep}{0.25cm}
\setlength{\jot}{10pt}
\titlespacing*{\section}{0pt}{4pt}{4pt}
\titlespacing*{\subsection}{0pt}{15pt}{1pt}

\fancyfoot{}
\fancyfoot[LO,RE]{\vspace{-7.1pt}\includegraphics[height=9pt]{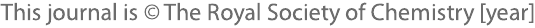}}
\fancyfoot[CO]{\vspace{-7.1pt}\hspace{13.2cm}\includegraphics{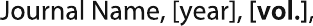}}
\fancyfoot[CE]{\vspace{-7.2pt}\hspace{-14.2cm}\includegraphics{head_foot/RF}}
\fancyfoot[RO]{\footnotesize{\sffamily{1--\pageref{LastPage} ~\textbar  \hspace{2pt}\thepage}}}
\fancyfoot[LE]{\footnotesize{\sffamily{\thepage~\textbar\hspace{3.45cm} 1--\pageref{LastPage}}}}
\fancyhead{}
\renewcommand{\headrulewidth}{0pt} 
\renewcommand{\footrulewidth}{0pt}
\setlength{\arrayrulewidth}{1pt}
\setlength{\columnsep}{6.5mm}
\setlength\bibsep{1pt}

\makeatletter 
\newlength{\figrulesep} 
\setlength{\figrulesep}{0.5\textfloatsep} 

\newcommand{\topfigrule}{\vspace*{-1pt}%
\noindent{\color{cream}\rule[-\figrulesep]{\columnwidth}{1.5pt}} }

\newcommand{\botfigrule}{\vspace*{-2pt}%
\noindent{\color{cream}\rule[\figrulesep]{\columnwidth}{1.5pt}} }

\newcommand{\dblfigrule}{\vspace*{-1pt}%
\noindent{\color{cream}\rule[-\figrulesep]{\textwidth}{1.5pt}} }

\makeatother

\twocolumn[
  \begin{@twocolumnfalse}
\vspace{3cm}
\sffamily
\begin{tabular}{m{4.5cm} p{13.5cm} }

\includegraphics{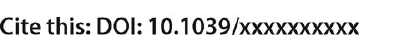} & \noindent\LARGE{\textbf{Electro-osmosis at surfactant-laden liquid-gas interfaces: beyond standard models}} \\ 
\vspace{0.3cm} & \vspace{0.3cm} \\

 & \noindent\large{Alexia Barbosa de Lima,\textit{$^{a}$} and Laurent Joly,$^{\ast}$\textit{$^{a}$}} \\

\includegraphics{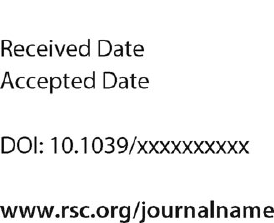} & \noindent\normalsize{Electro-osmosis (EO) is a powerful tool to manipulate liquids in micro and nanofluidic systems. While EO has been studied extensively at liquid-solid interfaces, the case of liquid-vapor interfaces, found e.g. in foam films and bubbles, remains to be explored. Here we perform molecular dynamics (MD) simulations of EO in a film of aqueous electrolyte covered with fluid layers of ionic surfactants and surrounded by gas. Following the experimental procedure, we compute the zeta potential from the EO velocity, defined as the velocity difference between the middle of the liquid film and the surrounding gas. We show that the zeta potential can be smaller or larger than existing predictions depending on the surfactant coverage. We explain the failure of previous descriptions by the fact that surfactants and bound ions move as rigid bodies and do not transmit the electric driving force to the liquid locally. Considering the reciprocal streaming current effect, we then develop an extended model, which can be used to predict the experimental zeta potential of surfactant-laden liquid-gas interfaces. 
%
} \\

\end{tabular}

 \end{@twocolumnfalse} \vspace{0.6cm}

  ]

\renewcommand*\rmdefault{bch}\normalfont\upshape
\rmfamily
\section*{}
\vspace{-1cm}


\footnotetext{\textit{$^{a}$~Univ Lyon, Universit\'e Claude Bernard Lyon 1, CNRS, Institut Lumi\`ere Mati\`ere, F-69622, LYON, France. E-mail: laurent.joly@univ-lyon1.fr}}




\section{Introduction}

Electrokinetic (EK) flows \new{(also called osmotic or surface-driven flows) are plug flows generated at interfaces by non-hydrodynamic forcing (e.g. electrical, thermal, or osmotic gradients). They represent} a powerful alternative to pressure-driven flows in micro and nanofluidic systems \cite{Anderson1989,Stone2004,Schoch2008,Bocquet2010}. In particular, electro-osmosis (EO), the flow induced by an electric field, is widely used to manipulate liquids in solid-state micro and nanofluidic devices. EO arises from the coupling between electrostatics and hydrodynamics close to charged interfaces, in the electrical double layer (EDL), a diffuse charged region in the liquid screening the surface charge \cite{Hunter2001}. The width of the EDL, called Debye length and denoted $\debye$, 
is typically nanometric in aqueous salt solutions \cite{Hunter2001}, so that EO is very sensitive to interfacial hydrodynamics \cite{Joly2004,Botan2013}. 

The amplitude of EO is quantified by the so-called zeta potential denoted $\zeta$, which is experimentally inferred from the measurement of the electro-osmotic velocity $v_\text{eo}$ \new{(the difference between the bulk liquid velocity -- outside the EDL -- and the wall velocity)} induced by an electric field $E$, \new{using the following expression, usually referred to as the (Helmholtz-)Smoluchowski equation} \cite{Delgado2007,Hunter2001}: 
\begin{equation}\label{eq:zetaEO}
v_\text{eo} = -\frac{\varepsilon \zeta}{\eta} E, 
\end{equation}
with $\varepsilon$ and $\eta$ the bulk liquid permittivity and shear viscosity, respectively. \new{According to the way it is experimentally measured}, the zeta potential is therefore a response coefficient, a priori controlled by both electrostatic and hydrodynamic properties of the interface \cite{Joly2004,Botan2013}. 
\new{Yet} on standard solid surfaces, by combining the Poisson equation for the electric potential (assuming a homogeneous permittivity of the liquid) and the Stokes equation for the velocity (assuming a homogeneous viscosity and a no-slip boundary condition applying slightly within the liquid), 
one can identify the zeta potential with the value of the electrical potential at the so-called shear plane, where the liquid velocity vanishes \cite{Hunter2001}. 
As the stagnant liquid layer between the solid surface and the shear plane is very thin (typically a fraction of nanometer), the zeta potential is usually a good (lower) estimate of the surface potential. 
Finally, the surface potential is commonly related to the surface charge using the Gouy-Chapman framework to describe the ion distribution in the EDL \cite{Hunter2001}. 
To sum up, the amplitude of EO flows is usually considered to be directly controlled by the surface potential, itself related to the surface charge. 

However this standard picture has been increasingly challenged in the past years \cite{Rotenberg2013,Knecht2013,Predota2016,Siboulet2017}. First, the no-slip boundary condition (BC) can fail at the nanoscale, and has to be replaced by a so-called partial slip BC first introduced by Navier \cite{Bocquet2007}. The partial slip BC relates the slip velocity $v_\text{s}$, i.e. the tangential velocity jump at the interface, to the shear rate $\dot{\gamma} = \partial v / \partial z$ (with $v$ the tangential velocity and $z$ the normal to the interface) in the liquid close to the interface: $v_\text{s} = b \dot{\gamma}$. 
In this relation $b$ is the so-called slip length, which can be interpreted as the depth within the solid surface where the linear extrapolation of the velocity profile vanishes. 
Typical slip lengths of water on smooth hydrophobic surfaces can reach tens of nanometers \cite{Bocquet2007}, so that the dynamics of the nanometric EDL should be strongly affected by slip. In fact, 
it was suggested that slip could enhance EK effects, with an amplification factor depending on the ratio between the slip length and the Debye length \cite{Muller1986,Stone2004}. This was later confirmed both numerically \cite{Joly2004,Joly2006} and experimentally \cite{Bouzigues2008,Audry2010}, highlighting that the zeta potential is not always representative of the surface potential and can be (much) larger. The viscosity and dielectric permittivity can also vary in the nanometric EDL, affecting the zeta potential \cite{Bonthuis2013,Majumder2015,Uematsu2017}. Additionally, Huang et al. \cite{Huang2007,Huang2008} have shown that ionic specificity toward hydrophobic surfaces could lead to electro-osmotic flows (hence finite zeta potentials) at uncharged surfaces, even though the adsorbed ions remained fully mobile. In that particular case, it was also shown that the zeta potential was the same for any finite slip length of the liquid on the hydrophobic surface. 
Finally, in a more recent work, the role of the mobility of the surface charge was examined \cite{Maduar2015}. Two limiting cases can be identified: if the surface charge is fixed, the standard description of EK effects applies, possibly corrected for liquid-solid slip or beyond-continuum effects. If on the other hand the surface charge is fully mobile (e.g. adsorbed charged species that remain mobile at the surface), the EK response can be described by the model of Huang et al. \cite{Huang2007,Huang2008}, where all the charged species are included in the liquid, and the solid surface is considered neutral.

While EK effects have been studied extensively at liquid-solid interfaces, much less is known on liquid-gas interfaces. 
Yet surfactant-laden liquid-gas interfaces can be found in systems with important industrial applications, e.g. liquid foams and bubbles \cite{Cantat,Kyzas2016,Huerre2014}.  
Applications of liquid foams are often limited by their instability and their inhomogeneity due to gravity-induced drainage \cite{Cantat,rio2014}. EK flows could then be used to control the liquid fraction and stability of foams \cite{Bonhomme2013,chevallier2013light,Fameau2011,miralles2014foam}. Liquid foams and foam films could also be used as low-cost micro and nanofluidic devices, using EK flows to drive liquids through the systems \cite{Bonhomme2013}. 
From a fundamental point of view also, the surface charge of the water-air interface is the object of hot debates \cite{Jungwirth2006,Vacha2007}. It seems therefore important to understand the relation between the zeta potential obtained from electrokinetic measurements, and the surface charge of these interfaces. 
Overall, it is therefore crucial to understand EK flows at surfactant-laden liquid-air interfaces in order to control and optimize the performance of such soft nanofluidic systems. 

However surfactant-laden liquid-gas interfaces are more complex than liquid-solid interfaces because the surface charge is carried by ionic surfactants, which penetrate inside the liquid and are mobile \cite{Langevin2014}. Moreover, the liquid+surfactants system is expected to strongly slip on the surrounding gas. It is therefore not clear whether the traditional model in terms of shear plane can be applied at this interface. A previous work using molecular dynamics simulations showed the existence of anomalous EK effects in foam films \cite{Joly2014}: this work observed a plateau of zeta potential for vanishing surfactant coverage, which was explained by the hydrodynamic slip of the liquid on the surfactant layer. At large surfactant coverage, a collapse of the zeta potential was explained by ion binding \cite{tarek1995,Kalinin:1996,bruce2002,Joly2014,Phan2016}, which canceled a fraction of the surface charge. However, \new{while in experiments the EO velocity is defined as the liquid velocity -- in the middle of the film -- with regard to the 
laboratory reference frame,} this work only considered the relative motion between the liquid and the surfactant layer, leaving aside the question of the dynamics of the surfactant layer \new{in the laboratory reference frame}. 

From this point of view, \new{surfactants can display a rigid or fluid behavior, depending on the surfactant nature and experimental conditions} \cite{Langevin2014}. \new{A so-called ``rigid'' surfactant layer behaves as a rigid wall}, remaining immobile in the reference frame of the laboratory, by creating a gradient of surface coverage and a related Marangoni stress. This corresponds typically to surfactants whose adsorption/desorption dynamics is slow as compared to the experimental timescale. 
In the case of \new{a rigid surfactant layer}, the previous numerical work considering the relative motion between the liquid and the surfactant layer \cite{Joly2014} also predicts the macroscopic motion of the liquid. 
On the other hand, \new{in a ``fluid'' surfactant layer, the} surfactants cannot build a gradient of surface concentration (e.g., if their adsorption/desorption dynamics is fast as compared to the timescale of the experiment); in that case, the surfactants will follow the liquid flow. This is the case we will consider here. \new{Since a fluid surfactant layer is free to move relatively to the laboratory reference frame, we need to also model the surrounding gas and to consider the relative motion between the liquid and the gas.} 



In this work, we performed molecular dynamics simulations of electro-osmosis in a foam film surrounded by a model gas. We computed the zeta potential from the EO velocity, defined as the relative velocity between the bulk liquid in the foam film and the gas \new{away} from the interface. We show that the zeta potential can be larger or smaller than standard predictions, depending on the surfactant coverage. We then trace back the failure of the standard models to the fact that surfactants and bound ions move as rigid bodies and do not transmit the electric driving force to the liquid locally. Considering the reciprocal streaming current effect, we then develop an extended model, \new{encompassed in Eq. \eqref{eq:zeta_new},} which can be used to predict the zeta potential in experiments. 

\section{Systems and methods}

%
\begin{figure}
\centering\includegraphics[width=0.8\linewidth]{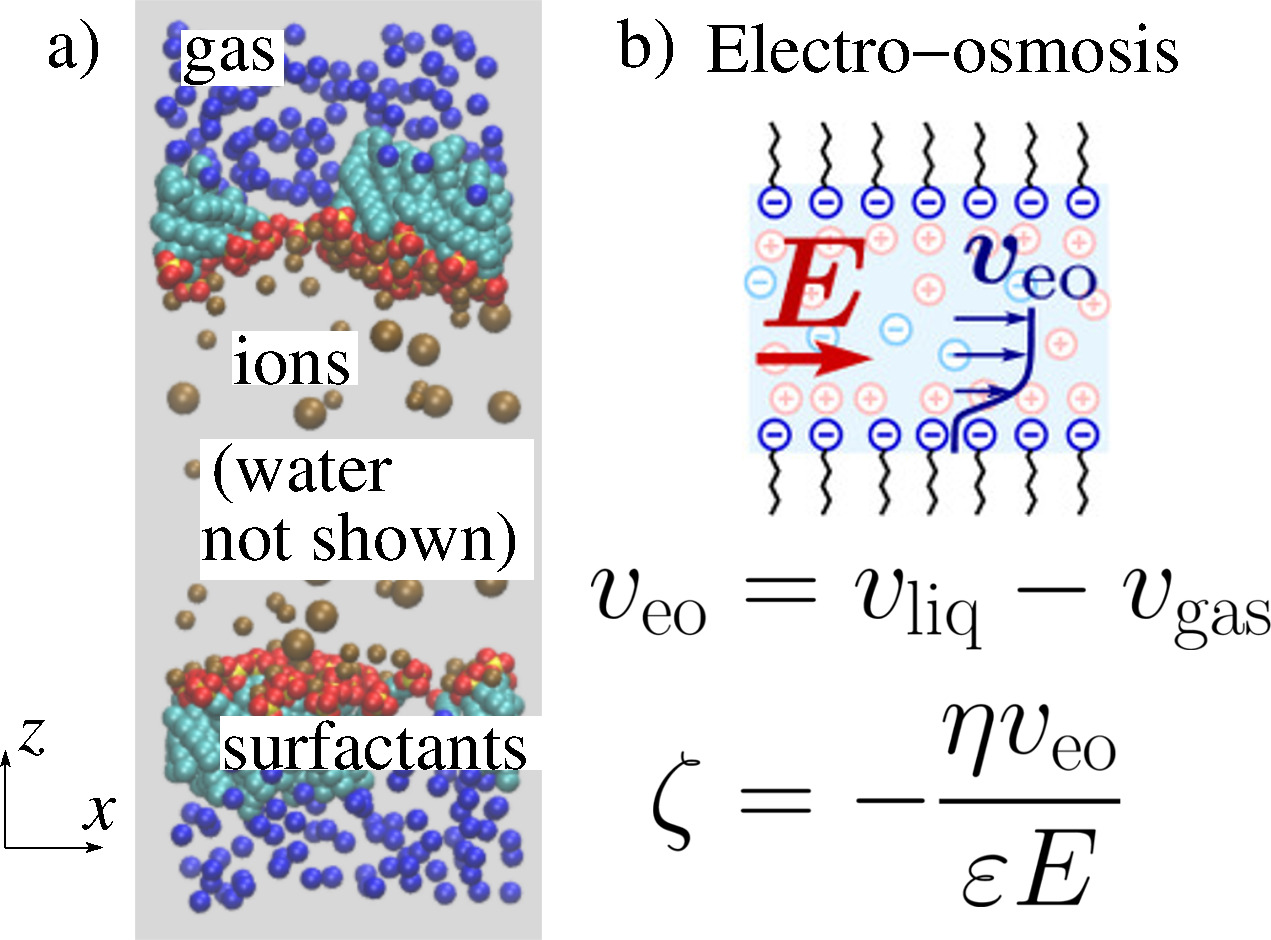}
\caption{a) Snapshot of a typical system (salt concentration $c_\text{s} = 0.26$\,M, surfactant coverage $c=1.5$\,nm$^{-2}$); gas atoms in blue, sodium and chloride ions in maroon, surfactant atoms: carbon in light blue, sulfur in yellow, oxygen in red; water molecules are not represented. b) Sketch of an electro-osmosis (EO) numerical experiment in the foam film, and formulas used to compute the zeta potential.} 
\label{fig:method}
\end{figure}
%

The numerical setup was based on the system used in Ref. \citenum{Joly2014}. In particular, 
we considered films of salty water (sodium chloride) coated with sodium dodecyl
sulfate (SDS) surfactants, see Fig. \ref{fig:method}. 
We used periodic boundary conditions 
\new{in all directions. In particular, periodic boundary conditions precluded the appearance of a gradient of surfactant coverage and a related Marangoni stress. We were therefore reproducing the experimental case of a so-called fluid surfactant layer. 
In the film plane, the box dimensions were} $L_x = L_y = 4.6$\,nm. 
We modeled water molecules, sodium ions and dodecyl
sulfate surfactants 
following Bresme and Faraudo \cite{Bresme2004,Bresme2006}. In
particular, we used the SPC/E (extended simple point charge) model of water, which provides reasonable values of both viscosity and dielectric constant. Chloride ions, not present in 
Refs.~\citenum{Bresme2004,Bresme2006}, were modeled consistently with
sodium ions, using the parameters of Dang~\cite{Dang1995}.
\new{We needed to keep the system size in the $z$ direction on the order of a few nanometers in order to be able to perform a large number of MD simulations on the system. To that aim, we chose a rather large salt concentration} $c_\text{s} = 0.26$\,M, with corresponding Debye length $\debye = 0.57$\,nm. The height of the foam films, along the $z$ direction, was fixed
to $L_z = 10\lambda_D = 5.7$\,nm, in order to prevent EDL overlap. 
For each salt concentration, the surface
density of surfactants $c$ (hereafter referred to as surfactant coverage) was varied 
by changing the number of surfactants in the unit cell, from 64 on each surface to 1, with a corresponding
surfactant coverage $c$ ranging 
from $0.047$ to $3.0$\,nm$^{-2}$, and a corresponding 
surface charge $\Sigma$ (assuming the surfactants are fully dissociated from their counter-ions) ranging from $-7.6$ to $-480$\,mC/m$^2$. 
As a comparison, surface densities up to 2.2\,nm$^{-2}$ have been
measured experimentally in the absence of salt~\cite{Bergeron:1997}.
\new{Note that in the simulations surfactants are only at the interface and there are no surfactants in the volume. Indeed, while in an experimental system there will be a dynamical adsorption/desorption equilibrium relating the bulk concentration and surface coverage of surfactants, the characteristic timescales are beyond the simulation times used here. Therefore, we chose to impose directly the surface coverage in the simulations. In experimental systems it is usually the bulk concentration that is controlled, but the surface coverage can be either measured or estimated using standard adsorption isotherms.} 

In this work we would like to explore the electro-osmotic motion of the liquid film with regards to the surrounding environment. Therefore, as a key new ingredient with regard to Ref. \citenum{Joly2014}, we added a gas surrounding the foam film. 
We added a layer of gas of ca. $1.5$\,nm on each side of the film, resulting in a ca. 3\,nm-thick gas layer over the periodic boundary. 
We considered a model gas made of particles interacting with all other particles through a repulsive Weeks-Chandler-Andersen potential, with parameters $\varepsilon = 0.15$\,kcal/mol and $\sigma = 3.8$\,\AA. 
We used a rather large density, so as to be able to measure the gas velocity with sufficient accuracy. We therefore made preliminary tests on the influence of the 
gas density on the results. For reasons that will appear later, we \new{did not} observe any effect of the gas density up to very dense gases. We consequently fixed the gas density to a value of ca. 
$1.6$\,nm$^{-3}$, well below the critical value where the results depended on the density, and large enough to obtain a sufficient statistics on the measurements.

The simulations were performed using LAMMPS \cite{lammps}. The initial
configurations were prepared using Packmol \cite{packmol}, and
the configuration files formatted for LAMMPS using the VMD \cite{vmd} TopoTools
plugin. 
The system was maintained at a temperature $T = 300$\,K using a
Nos\'e-Hoover thermostat, with damping time $200$\,fs, applied only to the directions perpendicular
to the flow. 
Long-range Coulombic interactions were computed using the
particle-particle particle-mesh (PPPM) method, and water molecules
were held rigid using the SHAKE algorithm. 
The equations of motion were solved using the velocity Verlet algorithm
with a timestep of $2$\,fs. After equilibration, on a relaxation time
of 1 to 2\,ns consistent with Ref. \citenum{Bresme2004}, production runs
lasted typically 5\,ns.

Two types of numerical experiments have been performed on these systems: 
electro-osmosis (EO) and streaming current (SC). 
In the EO configuration, an electric field is applied in
the $x$ direction, and the resulting EO flow is measured (Fig. \ref{fig:method}).
We fixed the applied electric field to $E = 0.2$\,V\,nm$^{-1}$. \new{This rather large value is not uncommon in non equilibrium MD \cite{Huang2007,Huang2008}, and is required in order to extract the output signal from thermal noise. Moreover, we tested that the system response remained linear up to a larger value of $E = 0.4$\,V\,nm$^{-1}$, so that the response coefficient (here the zeta potential) can be extrapolated to the much lower experimental values of the electric field.} 
We considered the relative motion between the liquid and the surrounding gas to
compute the EO velocity $v_{eo}$ in the middle of the
film and obtained  the zeta potential from Eq. \eqref{eq:zetaEO}. In that equation, \new{following the experimental procedure}, 
we used the tabulated values of the \new{bulk} viscosity and dielectric constant 
of SPC/E water at 300\,K, 
$\eta = 0.72$\,mPa\,s~\cite{Wu2006} and $\varepsilon_\text{r} = 70$~\cite{Reddy1989,Bonthuis2013}. 
\new{Indeed, although previous work showed that confinement could affect viscosity \cite{Raviv2001,Li2007} and permittivity \cite{Zhu2012,Zhang2013a}, the deviations appear generally for confinements stronger than the ones used here \cite{Markesteijn2012}, where the liquid film height is on the order of 20 water molecule size.  
%
Moreover, we would like to emphasize that we used the same values of $\eta$ and $\varepsilon$ both to deduce the zeta potential from the measured EO velocity and to estimate the theoretical zeta potentials, so that they are consistently affected by related errors and the comparison between theoretical models and numerical results should not be affected.   
Finally, the viscosity and permittivity of the liquid could be affected by the large salt concentration we used, but here also previous work indicated that the effect should remain small \cite{Marcus2009,Goldsack1977,Renou2014a}.} 

%
\begin{figure}
\centering\includegraphics[width=0.9\linewidth]{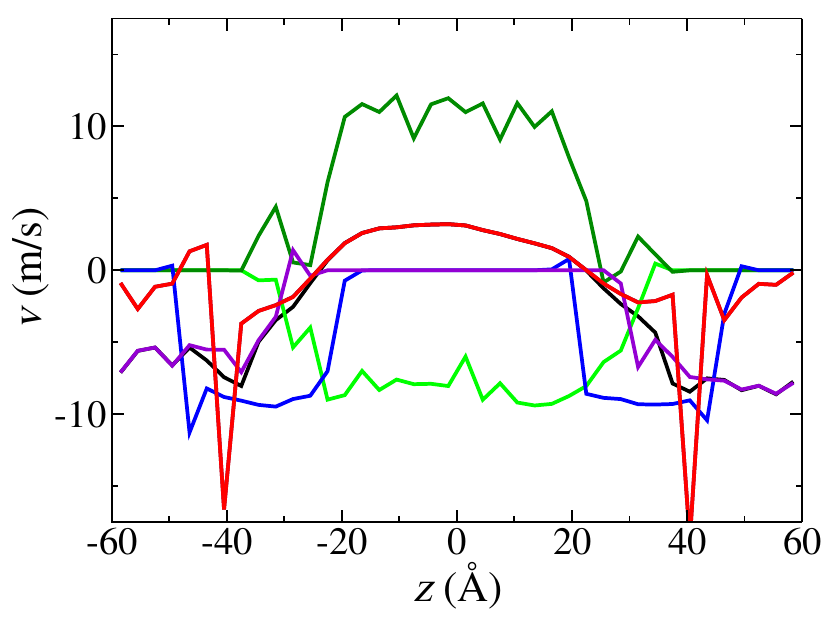}
\caption{Typical velocity profiles measured in an EO simulation (surfactant coverage $c=0.76$\,nm$^{-2}$). Black: all atoms; Red: liquid; Light green: chloride ions; dark green: sodium ions; blue: surfactants; violet: gas. \new{Note that the velocity profile for all atoms (in black) is hidden behind the liquid and gas velocity profiles in the liquid and gas region respectively, and briefly appears at the interfaces.}} 
\label{fig:EOvel}
\end{figure}
%

The simulation setup is isolated and globally neutral, so that the electric field introduces no net momentum. Therefore, both the liquid film, the surfactant layer and the gas region surrounding the surfactant-laden film are put into motion. In a macroscopic experimental setup, the gas will inevitably be in contact with a reservoir of momentum \new{-- immobile in the laboratory reference frame, so that the gas in contact with the foam film will be immobile in the steady state. 
Therefore the experimental EO velocity (given by the liquid velocity relative to the laboratory reference frame) will be adequately estimated in the simulations by computing the velocity difference between the liquid in the middle of the film and the surrounding gas.} 

We also conducted complementary SC simulations (see Sec. \ref{sec:discussion}), where a Poiseuille flow is generated in the $x$ direction, inducing an electric current due to the convective motion of the EDL. 
To generate the flow, a gravitylike force, adding up to $F$,  was applied to the liquid atoms, 
and a counterforce adding up to $-F$ was applied to the gas atoms. After testing the linear response of the system, we used a force $F = 1$\,kcal/mol/\AA\ for all the SC simulations. 


For the EO and SC configurations, we ran respectively 5 and 3 independent simulations from distinct 
initial configurations, in order to reduce statistical
uncertainties.

\section{Results}

Figure \ref{fig:EOvel} presents typical velocity profiles obtained in an EO simulation. As a result of the applied electric field, the different parts of the system are driven into motion: in particular, the liquid acquires a velocity relative to the surrounding gas. Ions and (anionic) surfactants display an electrophoretic motion, which superimposes to the motion of the solvent. From these velocity profiles, the electro-osmotic velocity $v_\text{eo}$ is computed as the difference between the liquid velocity in the middle of the foam film, $v_\text{liq}$, and the gas velocity away from the interface, $v_\text{gas}$: $v_\text{eo} = v_\text{liq} - v_\text{gas}$. The zeta potential is then computed using Eq. \eqref{eq:zetaEO}. 

%
\begin{figure}
\centering\includegraphics[width=0.9\linewidth]{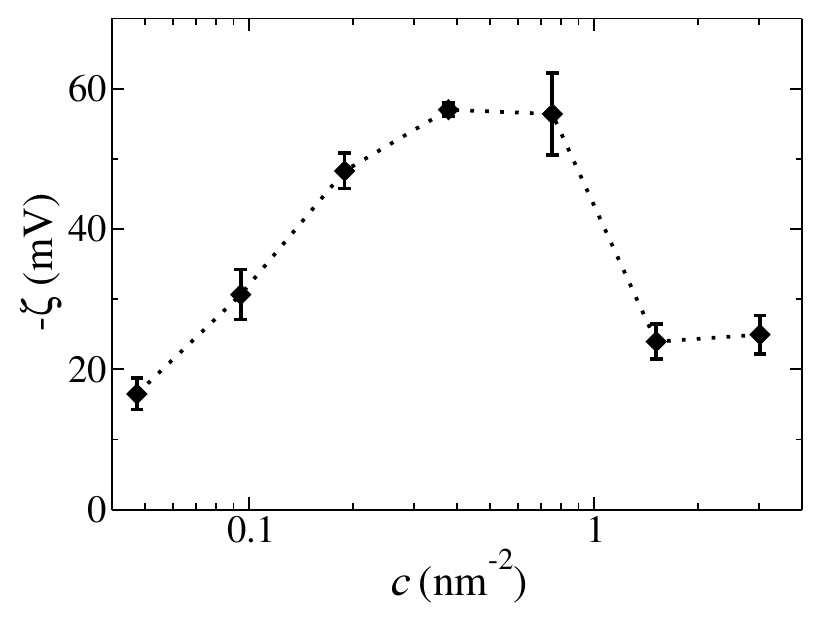}
\caption{Zeta potential as a function of surfactant coverage $c$, measured from EO simulations.} 
\label{fig:zeta}
\end{figure}
%

Figure \ref{fig:zeta} presents the evolution of the zeta potential with the surfactant coverage, which is shaped like a bell: the zeta potential first increases with the surfactant coverage, reaches a maximum of ca. $\sim 60$\,mV for a coverage of $c \sim 0.5$\,nm$^{-2}$, and \new{then collapses down to a value of ca.} 25\,mV. These results are in contrast with those obtained in a previous work \cite{Joly2014}, which considered the zeta potential related to the electro-osmotic motion of the liquid with regard to the surfactant layer. In particular we do not observe a plateau of the zeta potential at low surface coverage here. As discussed in the introduction, the previous work described the situation of a rigid surfactant layer --immobile in the reference frame of the laboratory, while the present simulations describe the case of a fluid surfactant layer. Therefore, it seems that the plateau of zeta potential (hence of the electro-osmotic response) can only be observed if the surfactant layer remains rigid even at low surface coverage. Interestingly, such a plateau has been observed experimentally \cite{Sasaki1991,Blanc2016}, suggesting that this is the case. It is in any case quite possible since the Marangoni stress ensuring the rigidity of the surfactant layer depends on the gradient of surface coverage and not on its absolute value \cite{Langevin2014}. 

In the following we will try to model the evolution of the zeta potential observed numerically in Fig. \ref{fig:zeta}. 

\subsection{Comparison with standard models}

\subsubsection{Ideal surface potential}

\begin{figure}
\centering\includegraphics[width=0.9\linewidth]{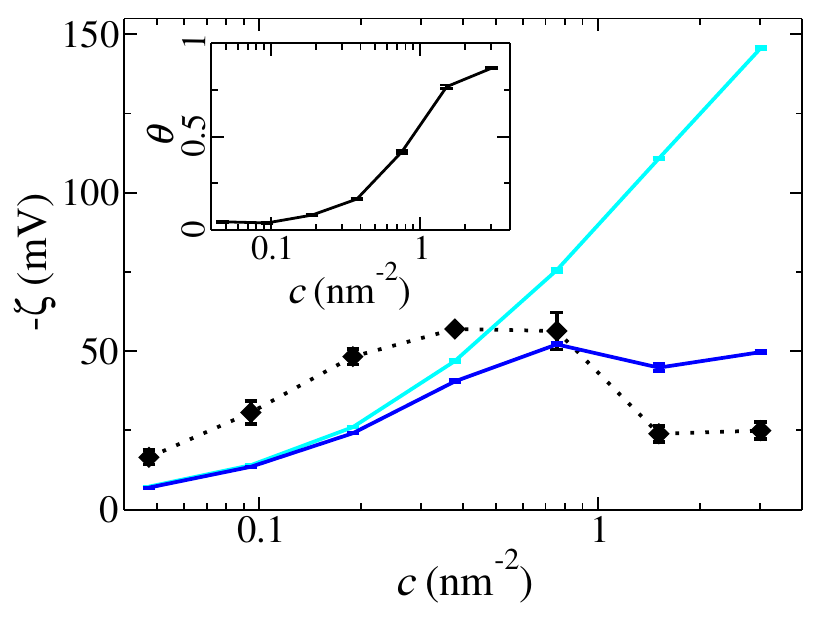}
\caption{Zeta potential as a function of surfactant coverage $c$: comparison between the numerical results (black diamonds and dotted line) and the ideal surface potential model (see text for details), assuming full dissociation of the ionic surfactants (cyan line) or taking into account ion binding (blue line)\new{, using data presented in the inset}. Inset: fraction of surfactant charge canceled by ion binding $\theta$ as a function of surfactant coverage $c$, measured using a geometric criterion (see text for details).} 
\label{fig:V0}
\end{figure}

As discussed in the introduction, the zeta potential is traditionally identified with the electrical potential at the shear plane, and the latter is considered a good (lower) estimate of the surface potential \cite{Hunter2001}. We therefore started by comparing the measured zeta potential values with the surface potential $V_0$ of an infinitely thin surface bearing the same surface charge $\Sigma$ as the surfactant layer, obtained within the Poisson-Boltzmann framework (i.e., neglecting correlations and non-electrostatic interactions) \cite{Hunter2001}: 
\begin{equation}\label{eq:V0}
V_0 = \frac{2\kt}{e} \mathrm{asinh} \left( \frac{\Sigma}{4 e c_\text{s} \debye} \right), 
\end{equation}
with $k_\text{B}$ the Boltzmann constant, $T$ the temperature, $e$ the elementary charge, and $c_\text{s}$ the bulk salt concentration. We will refer to $V_0$ as the ideal surface potential because Eq. \eqref{eq:V0} ignores the fact that the surfactant charge is distributed in the liquid and overlaps with the ion charge. 
We first assumed that the surfactants were fully dissociated from their counterion, with each surfactant head carrying a negative elementary charge $-e$, so that the surface charge was proportional to the surfactant coverage: $\Sigma = -e c$. Figure \ref{fig:V0} shows that this first model underestimates the zeta potential at low surfactant coverage, and strongly overestimates it at high coverage. The low coverage limit is unexpected since in the standard picture, the zeta potential should always be a lower estimate of the surface potential. At high coverages, it has been discussed in previous work that ion binding will cancel a fraction of the surface charge \cite{tarek1995,Kalinin:1996,bruce2002,Joly2014,Phan2016}. We computed the fraction of bound ions $\theta$ using a geometrical criterion detailed in the supplemental material of Ref. \citenum{Joly2014}, see $\theta$ versus $c$ in the inset of Fig. \ref{fig:V0}, and recomputed $V_0$ accounting for ion binding in the surface charge: $\Sigma_\text{ib} = -e(1-\theta) c$, see Fig. \ref{fig:V0}. Taking ion binding into account does not change the value of the surface potential at low surfactant coverage, because the fraction of bound ions is very low. At large surfactant coverage, taking into account ion binding brings $V_0$ closer to the measured zeta potential, but the values still \new{do not} match. 
In conclusion, a simple estimate of the surface potential underestimates the zeta potential at low surfactant coverage, and overestimates it at high coverage, even when accounting for ion binding occurring at large coverage. We therefore turned to a more sophisticated model introduced by Huang et al. \cite{Huang2007,Huang2008}. 

%
%

\subsubsection{Model of Huang et al.}

Huang et al. modeled the EO flow of a liquid in contact with a slipping hydrophobic surface under an electric field $E$ parallel to the interface \cite{Huang2007,Huang2008}. They solved Stokes equation: 
\begin{equation}\label{eq:Stokes}
- \eta \frac{\mathrm{d}^2 v}{\mathrm{d} z^2} = \rho_e(z) E, 
\end{equation}
with $z$ the direction normal to the interface, $v(z)$ the velocity along the direction of the applied electric field $E$, $\eta$ the shear viscosity of the liquid (assumed homogeneous), and $\rho_e(z)$ the charge density profile in the liquid. Assuming a partial slip BC with a slip length $b$, they computed the electro-osmotic velocity in the bulk liquid and the corresponding zeta potential: 
\begin{equation}\label{eq:Huang}
\zeta = - \frac{\eta v_\text{eo}}{\varepsilon E} = - \frac{1}{\varepsilon} \int_{z_\text{s}}^{z \gg \debye} (z - z_\text{s} + b) \rho_e(z) \, \mathrm{d}z, 
\end{equation}
where $\varepsilon$ is the bulk liquid permittivity, $z_\text{s}$ the shear plane position where the partial slip BC applies, and $z \gg \debye$ is located in the bulk liquid away from the EDL. 
\new{Note that Eq. \eqref{eq:Huang} does not make any assumption on the permittivity profile at the interface. Indeed, the model predicts the EO velocity based on Stokes equation only, without any modeling of the electrostatics.
The bulk liquid permittivity only appears in this equation because the zeta potential is deduced from the EO velocity using Eq. \eqref{eq:zetaEO}, following the experimental procedure.}

As discussed in Ref. \cite{Huang2007,Huang2008}, this expression can be simplified for an uncharged surface, i.e. when all the charges -- whether they are adsorbed or not -- remain fully mobile and can be considered as belonging to the liquid. Indeed, in that case the liquid is also globally uncharged, i.e. $\int \rho_e(z) \, \mathrm{d}z = 0$, so that the constant terms $z_\text{s}$ and $b$ multiplied by $\rho_e(z)$ in the integral of Eq. \eqref{eq:Huang} will disappear, and the integral can be run from within the uncharged surface, $z = - \infty$: 
\begin{equation}\label{eq:Huang_uncharged}
\zeta^{\Sigma=0} = - \frac{1}{\varepsilon} \int_{-\infty}^{z \gg \debye} z \, \rho_e(z) \, \mathrm{d}z. 
\end{equation}
Therefore, the zeta potential at an uncharged surface does not depend on the slip length. Indeed, if the liquid is globally neutral, the electric field does not apply any net force on the liquid, so that in the steady-state the net force between the liquid and the surface must vanish. 
\new{Consequently, the viscous shear stress applied by the liquid on the surface must vanish, $\eta\, \mathrm{d}v/\mathrm{d}z |_{z=\zs} = 0$, hence the shear rate must vanish too: $\mathrm{d}v/\mathrm{d}z |_{z=\zs} = 0$. The partial slip BC then imposes that the slip velocity also vanishes at the interface: $v_\text{s} = b\, \mathrm{d}v/\mathrm{d}z |_{z=\zs} = 0$. As a result, the boundary condition for the flow (zero velocity and zero shear rate) will be the same for any finite value of the slip length.} 

%
\begin{figure}
\centering\includegraphics[width=0.9\linewidth]{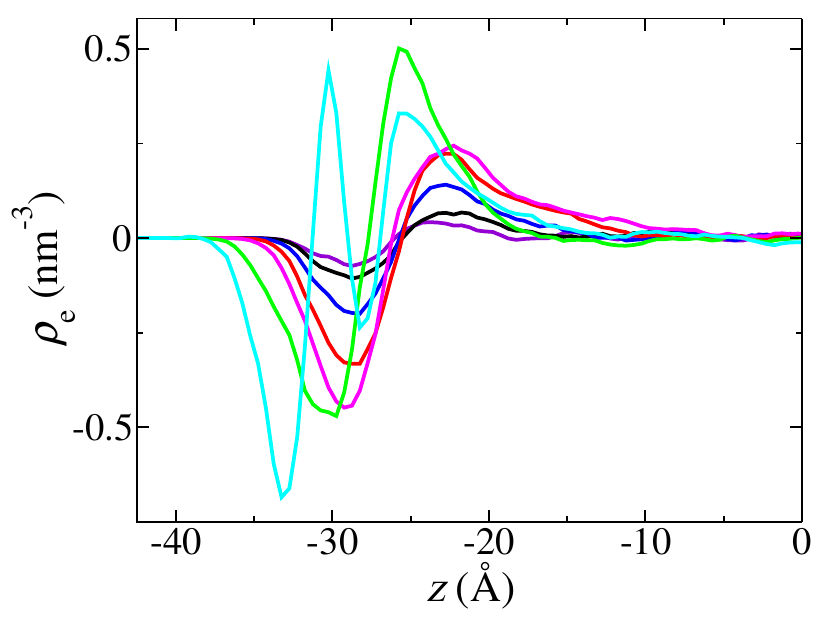}
\caption{Total charge density profiles at the surfactant-laden gas-liquid interface, for 7 different surfactant coverages from $0.047$ to $3.0$\,nm$^{-2}$ (by increasing coverage: violet, black, blue, red, magenta, green, cyan).} 
\label{fig:rhoe}
\end{figure}
%

\begin{figure}
\centering\includegraphics[width=0.9\linewidth]{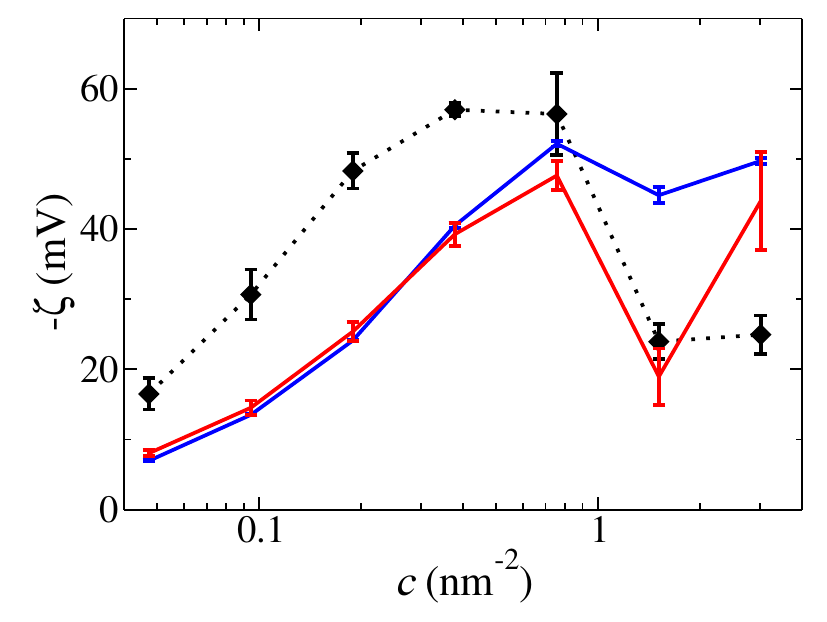}
\caption{Zeta potential as a function of surfactant coverage $c$: comparison between the numerical results (black diamonds and line), the ideal surface potential model with ion binding (blue line), and Huang et al.'s model (red line), see text for details.} 
\label{fig:Huang}
\end{figure}

The surfactant-laden liquid-gas interface we consider here is very similar to the case described by Huang et al. \cite{Huang2007,Huang2008}. The surfactants are fully mobile and can be included in the liquid part of the system, and the gas plays the role of a neutral, very slippery surface. We therefore compared our results and the ideal surface potential with the prediction of Huang et al.'s model for a neutral surface. To that aim, we measured the charge density profiles $\rho_e(z)$ in the simulations at different surfactant coverage $c$ (Fig. \ref{fig:rhoe}), and injected them in Eq. \eqref{eq:Huang_uncharged}, see Fig. \ref{fig:Huang}. At low surfactant coverage, Huang et al. model underestimates the zeta potential. Surprisingly, it matches almost perfectly the ideal surface potential prediction. At large surfactant coverage, Huang et al. prediction also fails to predict the measured zeta potential. This overall failure of the model is unexpected because this model is simply based on Stokes equation and makes very few assumptions \new{(essentially the viscosity is assumed homogeneous in the EDL)}. 
Furthermore, this model was able to quantitatively predict the zeta potential of hydrophobic surfaces with specific ion adsorption in previous work \cite{Huang2007,Huang2008}\new{, suggesting that in particular the assumption of a homogeneous viscosity is reasonable}. 

%
%

\section{Discussion}
\label{sec:discussion}

While the failure of the ideal surface potential to describe the zeta potential could be expected given the complexity of the surfactant-laden liquid-vapor interface, the fact that the more general model of Huang et al. also fails is more surprising. However, a first hint of the limit of Huang et al.'s model comes from the fact that Eqs. \eqref{eq:Huang} and \eqref{eq:Huang_uncharged} are based on the total charge density profile, so that they are unable to account for ion binding, where bound ions effectively neutralize the surface charge. Indeed, as it is written, the Stokes equation, Eq. \eqref{eq:Stokes}, assumes that the electrical force applied by the electric field on charged particles is transmitted directly and locally to the liquid. However this is not the case for the surfactants and bound ions, because these charges are rigidly bound together. 
\new{In the presence of spatially extended clusters of atoms (surfactants and possibly bound ions), the electric field applies an electric force to the clusters (depending on their total charge), and the clusters then apply a drag force to the liquid. In the steady state, the total electrical force experienced by a cluster is equal to the total drag force it applies on the liquid, but the spatial distribution of the forces are in general different (for instance, electrically neutral atoms belonging to a cluster will nevertheless apply a drag force to the liquid). Therefore, the external volumic force experienced by the liquid is not equal to the local electric force as assumed in Huang et al.'s model.}  
To illustrate this point, one can think of the extreme case where all counter-ions are bound to the surfactants: in that case, the rigid body formed by surfactant molecules and bound ions will be globally neutral, so that the electric field will apply no net force on this rigid body, which will consequently transmit no force to the liquid. 

\subsection{A detour through streaming current}
\label{sec:SC}

However, it is not trivial to describe the force redistribution in the EO configuration. Therefore, in order to extend the model of Huang et al. to account for the rigid motion of the surfactants and bound ions, we will now turn to the reciprocal streaming current (SC) configuration \cite{Hunter2001}.   
While in EO a gradient of electrical potential induces a hydrodynamic flow, SC refers to the electrical current induced by a gradient of pressure. Indeed, a pressure gradient creates a Poiseuille flow, which put the charged EDL into motion and generates a convective electrical current. According to Onsager reciprocal relations \cite{Brunet2004}, the SC is also quantified by the zeta potential: 
\begin{equation}\label{eq:Ie}
\frac{I_e}{A} = - \frac{\epsilon \zeta}{\eta} (-\nabla p), 
\end{equation}
with $A$ the cross-section of the channel. 

%
\begin{figure}
\centering\includegraphics[width=0.9\linewidth]{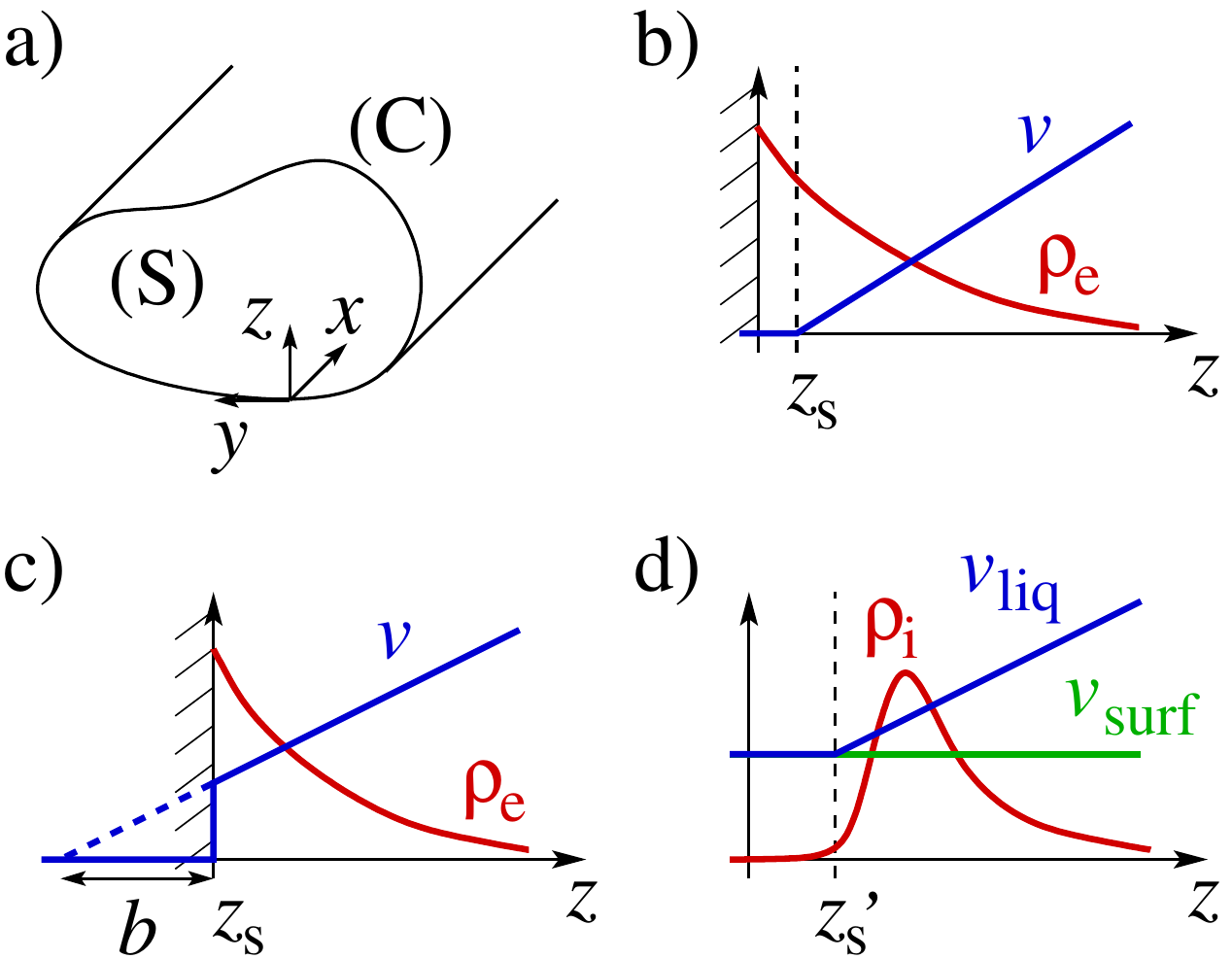}
\caption{a) Streaming current (SC) in a channel of arbitrary cross-section $\mathcal{S}$, with a thin EDL. 
The SC is the sum of infinitesimal currents $\mathrm{d}I_e$ generated at line elements $\mathrm{d}y$ of the contour $\mathcal{C}$ of the cross-section. These infinitesimal currents are the integrals along the local normal $z$ to the contour, of the electrical flux $j_\text{e} = \rho_\text{e} v$, with $\rho_\text{e}$ the charge density and $v$ the liquid velocity along the $x$ direction. 
b) Typical charge density and velocity profiles at a solid surface with a no-slip boundary condition (BC) applying at the shear plane $\zs$. The velocity profile writes $v = \dot{\gamma} (z-\zs)$ for $z>\zs$, where $\dot{\gamma}$ is the local shear rate close in the EDL. 
c) Same as b) but with a partial slip BC (slip length $b$). In that case the velocity profile writes: $v = \dot{\gamma} (z-\zs+b)$.
d) At a surfactant-laden gas-liquid interface, the electrical flux $j_\text{e}$ is given by the product of the free ion density profile $\rho_\text{i}$ and the excess velocity profile $v_\text{ex} = v_\text{liq} - v_\text{surf}$. The excess velocity profile writes: $v_\text{ex} = \dot{\gamma} (z-\zsp)$ for $z>\zsp$. } 
\label{fig:SC}
\end{figure}
%

Here again, assuming a homogeneous liquid permittivity and viscosity, and a no-slip BC applying at the shear plane $z_\text{s}$, the zeta potential can be related to the electrical potential at the shear plane \cite{Hunter2001}. To that aim, 
let's consider a channel of arbitrary cross-section $\mathcal{S}$ (see Fig. \ref{fig:SC}.a), and assume the Debye length is much smaller than the channel typical size. 
The SC $I_\text{e}$ originates from the convective motion of the EDL, located along the contour $\mathcal{C}$ of the channel cross-section. Defining a local reference frame along the contour, with $x$ the direction of the flow, $y$ and $z$ the local tangent and normal to the contour, respectively, $I_\text{e}$ can therefore be expressed as the sum of the infinitesimal currents $\mathrm{d}I_\text{e}$ arising at line elements $\mathrm{d}y$ of the contour (Fig. \ref{fig:SC}.b): 
\begin{equation}
I_\text{e} = \oint_\mathcal{C} \mathrm{d}I_\text{e},   
\end{equation}
with 
\begin{equation}
\mathrm{d}I_\text{e} = \mathrm{d}y \int_{z_\text{s}}^{z \gg \debye} \mathrm{d}z\ \rho_\text{e}(z)\  v(z) , 
\end{equation}
where $z$ is the local normal to the surface and $z_\text{s}$ the position of the shear plane. 
Here it is assumed that the charged particles have the same velocity as the flow, which is quite reasonable for small mobile ions.
\new{If the Debye length is much smaller than the channel size, the curved Poiseuille velocity profile can be approximated by a simple shear velocity profile over the extent of the EDL (with a no-slip BC applying at the shear plane $\zs$): $v(z)=(z-z_\text{s}) (\mathrm{d}v/\mathrm{d}z)|_{z=z_\text{s}}$ (Fig. \ref{fig:SC}.b).} 
Using Poisson equation $\rho_\text{e} (z) = -\varepsilon (\mathrm{d}^2 V / \mathrm{d}z^2)$ (assuming a constant $\varepsilon$), together with the boundary conditions $V(z \gg \debye)=0$ \new{and} $(\mathrm{d}V / \mathrm{d}z)|_{z \gg \debye}=0$,  
the infinitesimal current can be written: 
\begin{equation}
\mathrm{d}I_\text{e} = -\varepsilon V(z_\text{s}) \mathrm{d}y \left.\frac{\mathrm{d}v}{\mathrm{d}z}\right|_{z=z_\text{s}}, 
\end{equation}
so that: 
\begin{equation}
I_\text{e} = -\varepsilon V(z_\text{s}) \oint_\mathcal{C} \mathrm{d}y \left.\frac{\mathrm{d}v}{\mathrm{d}z}\right|_{z=z_\text{s}}.  
\end{equation}

The second Green theorem can be used to transform this integral over the contour of the channel into an integral over the cross-section of the channel of the Laplacian of the velocity: 
\begin{equation}
I_\text{e} = -\varepsilon V(z_\text{s}) \iint_\mathcal{S} \mathrm{d}S\ \nabla^2 v. 
\end{equation}
One can finally use Stokes equation, $\nabla^2 v = -\nabla p / \eta$ (assuming a constant viscosity), to obtain an expression for the SC as a function of the pressure gradient, 
\begin{equation}
I_\text{e} = -\varepsilon\ V(z_\text{s})\ A (-\nabla p / \eta), 
\end{equation}
and for the corresponding zeta potential: 
\begin{equation}
\zeta = \frac{\eta \nabla p}{\varepsilon A} I_\text{e} = V(z_\text{s}). 
\end{equation}

Incidentally, one can note a direct relation between the infinitesimal current and the zeta potential: 
\begin{equation}
\zeta = -\frac{\mathrm{d}I_\text{e}}{\varepsilon \mathrm{d}y \left.\frac{\mathrm{d}v}{\mathrm{d}z}\right|_{z=z_\text{s}}}. 
\end{equation}
Consequently, the formula of Huang et al., Eq. \eqref{eq:Huang}, simply describes the infinitesimal current related to a charge distribution $\rho_\text{e} (z)$ and a linear velocity profile with a partial slip BC, $v(z)=(z-z_\text{s}+b) (\mathrm{d}v/\mathrm{d}z)|_{z=z_\text{s}}$ (Fig. \ref{fig:SC}.c):
\begin{equation}
\mathrm{d}I_\text{e} = \mathrm{d}y \int_{z_\text{s}}^{z \gg \debye} \mathrm{d}z\ \rho_\text{e}(z)\  (z-z_\text{s}+b) (\mathrm{d}v/\mathrm{d}z)|_{z=z_\text{s}} ,
\end{equation}
so that: 
\begin{equation}
\zeta = -\frac{\mathrm{d}I_\text{e}}{\varepsilon \mathrm{d}y \left.\frac{\mathrm{d}v}{\mathrm{d}z}\right|_{z=z_\text{s}}} = - \frac{1}{\varepsilon} \int_{z_\text{s}}^{z \gg \debye} (z - z_\text{s} + b) \rho_e(z) \, \mathrm{d}z. 
\end{equation}

\subsection{Extending Huang et al.'s model}

As mentioned before, in the previous description it is assumed that the charged species in the liquid simply follow the flow, so that their velocity profile identifies with that of the liquid. This assumption works quite well for micro-ions, but the charged atoms in surfactant molecules and the bound ions form an extended rigid body, so that they must all flow with the same velocity (plug flow). In contrast, the liquid velocity varies over the extent of this rigid body, so that the plug flow of the charges in the rigid body and the liquid flow cannot match everywhere. 

%
\begin{figure}
\centering\includegraphics[width=0.9\linewidth]{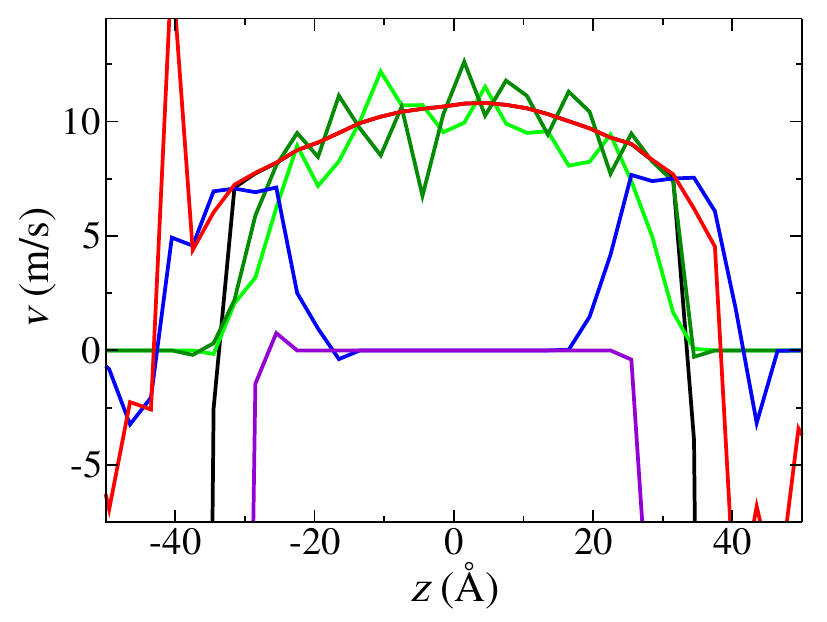}\\
\centering\includegraphics[width=0.9\linewidth]{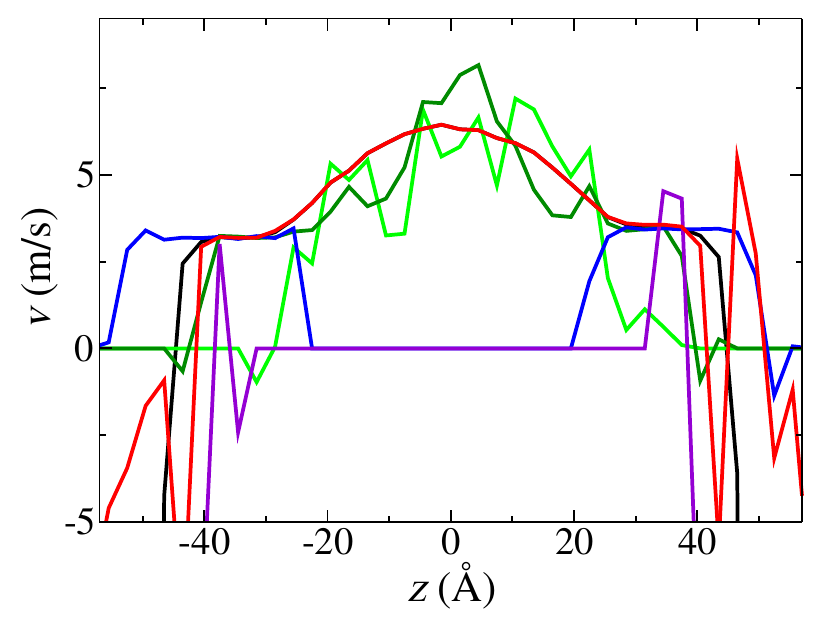}
\caption{Typical velocity profiles measured in a SC simulation, for surfactant coverages $c=0.19$\,nm$^{-2}$ (top) and $c=3.0$\,nm$^{-2}$ (bottom). Black: all atoms; Red: water; light green: chloride ions; dark green: sodium ions; blue: surfactants; violet: gas. The gas velocity is strongly negative, well beyond the limits of the figure.} 
\label{fig:SCvel}
\end{figure}
%

In order to illustrate this point, we performed complementary SC simulations. Figure \ref{fig:SCvel} represents typical flow profiles in SC simulations with two different surfactant coverage. In all cases, it can be seen that water and free micro-ions share the same parabolic velocity profile, and that they strongly slip on the surrounding gas. On the other hand, the surfactants and bound ions move with a homogeneous velocity, on the order of the water velocity at the liquid-gas interface. 
In order to derive an expression for the SC and for the related zeta potential, we decompose the water+free ions flow profile into two contributions: on the one hand, a plug flow profile with the velocity of the surfactants+bound ions, and on the other hand an excess flow profile vanishing at the ``effective shear plane'' $z_s'$ where the water+free ions velocity meets the surfactant+bound ions velocity. The overall plug flow motion of the entire foam film (i.e., water+free ions and surfactants+bound ions) at the velocity of the surfactants does not generate any current since the system is globally neutral. Therefore, only the excess velocity profile of water+free ions will generate a net current, which can be written: 
\begin{equation}
\mathrm{d}I_\text{e} = \mathrm{d}y \int_{z_\text{s}'}^{z \gg \debye} \mathrm{d}z\ \rho_\text{i}(z)\ v_\text{ex} (z) ,
\end{equation}
with $\rho_i (z)$ the charge density carried by free ions, and $v_\text{ex} (z) = (z-z_s') (\mathrm{d}v/\mathrm{d}z)|_{z=z_s'}$ the excess velocity profile, where the curvature of the \new{Poiseuille} velocity profile is neglected \new{over the extent} of the EDL (Fig. \ref{fig:SC}.d). One can then deduce a new formula for the zeta potential: 
\begin{equation}\label{eq:zeta_new}
\zeta = -\frac{\mathrm{d}I_\text{e}}{\varepsilon \mathrm{d}y \left.\frac{\mathrm{d}v}{\mathrm{d}z}\right|_{z=z_\text{s}}'} = - \frac{1}{\varepsilon} \int_{z_\text{s}'}^{z \gg \debye} (z - z_\text{s}') \rho_\text{i}(z) \, \mathrm{d}z. 
\end{equation}
This is the main prediction of this article, relating the zeta potential to the charge density profile of free ions and to the effective shear plane, where the water+free ions velocity meets the surfactant+bound ions velocity in the SC configuration. 
\new{Following Huang et al.'s original model, this prediction makes no assumption on the local dielectric properties of the liquid, since its basic ingredient is directly the free charge density profile (the bulk dielectric permittivity only appears in the formula because the zeta potential is deduced from the EO velocity using Eq. \eqref{eq:zetaEO}, according to the experimental procedure). Our model only assumes that the viscosity is homogeneous above $\zsp$, which is compatible with direct observation of the numerical velocity profiles in the SC current configuration, i.e. the Poiseuille excess velocity profiles are well described by a simple parabola starting at $\zsp$.} 
As a side note, one can now understand the negligible influence of the gas density on the zeta potential observed during our preliminary tests. Indeed, while the gas density will modify the friction of the surfactant-laden liquid film on the gas, this will only affect the relative motion between the film and the gas, which has no impact on the SC, hence on the zeta potential. 

%
\begin{figure}
\centering\includegraphics[width=0.9\linewidth]{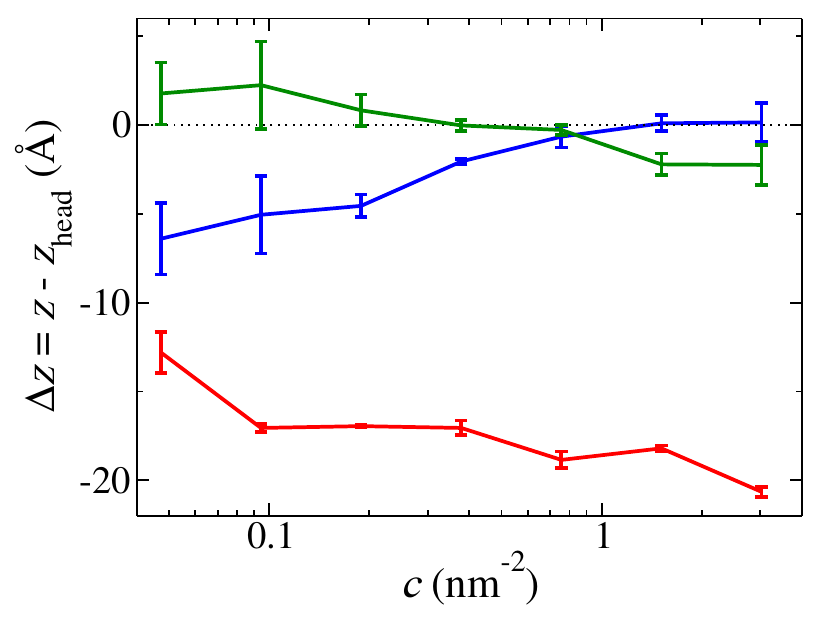}
\caption{Evolution of the position $z_\text{s}$ of the slip plane between the liquid -- water+ions+surfactants -- and the gas (red line), of the effective shear plane position $z_\text{s}'$ where the water+free ions and surfactants+bound ions velocity profiles meet (blue line), and of the effective electrostatic wall position $z_\text{ES}$ (green line), as a function of the surfactant coverage $c$. The relative values with regard to the average position of the sulfur atoms of the surfactant heads, $z_\text{head}$, are plotted: $\Delta z_\text{s} = z_\text{s} - z_\text{head}$, $\Delta z_\text{s}' = z_\text{s}' - z_\text{head}$, and $\Delta z_\text{ES} = z_\text{ES} - z_\text{head}$.} 
\label{fig:zs}
\end{figure}
%

Due to the large noise on the velocity profiles obtained from SC simulations, we chose not to extract the value of $\zsp$ from the SC velocity profiles, but instead we used $\zsp$ as a fitting parameter to reproduce the zeta potentials obtained from EO simulations. We were then able to reproduce perfectly the EO results, and we checked that the obtained values of $\zsp$ for varying surface coverage of surfactants were compatible with the SC velocity profiles. Figure \ref{fig:zs} shows the evolution of both the position $z_\text{s}$ of the slip plane between the liquid -- water+ions+surfactants -- and the gas, and the effective shear plane position $\zsp$. We first emphasize that $\zs$ and $\zsp$ are quite different, and that only $\zsp$ has an impact on the zeta potential, since the global slippage between the gas and the (neutral) liquid generates no current. The effective shear plane position $\zsp$ evolves smoothly from slightly within the surfactant layer (ca. 6\,\AA\ below the S atom of the surfactant head) to the top of the surfactant layer (i.e., at the position of the S atoms of the heads) with increasing surfactant coverage. In a previous work, we have rationalized this behavior based on a simple hydrodynamic model (see supplemental material of Ref. \citenum{Joly2014}). Here we take the measured evolution of $\zsp$ with $c$ as an input to the synthetic model we will propose later. 

Indeed, while Eq. \eqref{eq:zeta_new} can be used to predict the zeta potential once the distribution of free ions at the interface is known (e.g. by running relatively short equilibrium molecular dynamics simulations), 
we finally took an additional step to simplify the equation predicting the zeta potential of a liquid-vapor interface laden with fluid surfactants, in order to highlight the mechanisms underlying the unexpected discrepancy between the ideal surface potential and the zeta potential. 

%
\begin{figure}
\centering\includegraphics[width=0.9\linewidth]{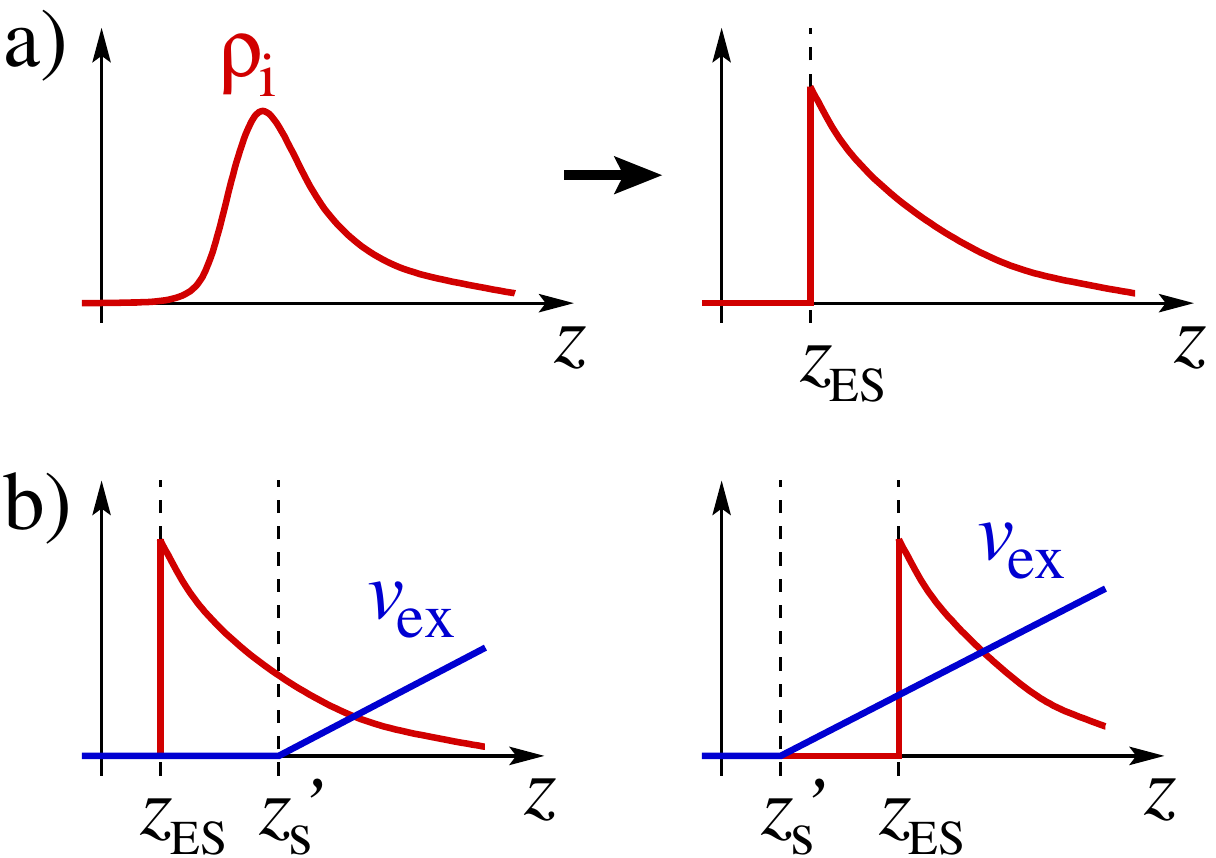}
\caption{a) Replacing the charge density profile of free ions by the charge density profile of an effective electrostatic surface located at $z=\zes$. 
b) Left: if $\zes<\zsp$, there is an effective stagnant layer of width $\zsp-\zes$; right: if $\zsp<\zes$, there is an effective partial slip BC with a slip length $b_\text{eff} = \zes - \zsp$.} 
\label{fig:zes}
\end{figure}
%

To that aim, we replace the measured free charge density profile by the effective ideal charge density profile generated by an infinitely thin charged surface located at a position $\zes$, within the Poisson-Boltzmann framework (Fig. \ref{fig:zes}.a). Combining Eq. \eqref{eq:zeta_new} and Poisson equation, the zeta potential can be expressed as a function of the electrostatic potential, following the same approach as in section \ref{sec:SC}. Two different cases can arise: first, if the effective shear plane is located more inside the liquid than the effective electrostatic surface (Fig. \ref{fig:zes}.b), then following the same approach as in section \ref{sec:SC}, the zeta potential is given by the ideal electrostatic potential at the effective shear plane, $\zeta = V(\zsp-\zes)$, using the general solution of the Poisson-Boltzmann equation for the ideal potential profile: 
\begin{equation}\label{eq:V_PB}
V(z) = \frac{4 \kt}{e} \text{atanh} \left[ \tanh \left( \frac{e V_0}{4 \kt}\right) \text{e}^{-\sfrac{z}{\debye}} \right]. 
\end{equation}
If on the contrary, the effective shear plane falls on the vapor side as compared to the effective electrostatic surface (Fig. \ref{fig:zes}.b), then the situation can formally be described as a slip boundary condition applying at the effective electrostatic surface, and with a slip length given by $b_\text{eff} = \zes-\zsp$. It has been shown in previous work \cite{Muller1986,Joly2004} that the zeta potential then writes: 
\begin{equation}\label{eq:zeta_beff}
\zeta = V_0 \left( 1 + \frac{b_\text{eff}}{\debye} \right). 
\end{equation}

We used Eqs. \eqref{eq:V_PB} and \eqref{eq:zeta_beff} to compute the position of the effective electrostatic surface in order to reproduce the zeta potential obtained from EO simulations. Figure \ref{fig:zs} compares the positions of $\zsp$ and $\zes$ for varying surfactant coverage. The effective ES surface shifts regularly from slightly above (2\,\AA) the sulfur atom of the surfactant heads at low coverage, to slightly below (2\,\AA) at high coverage. 

Interestingly, the comparison between $\zsp$ and $\zes$ in Fig. \ref{fig:zs} provides a simple interpretation for the discrepancy between the zeta potential and the ideal surface potential: at low surfactant coverage, the effective shear plane is located on the vapor side of the electrostatic surface, so that the zeta potential is amplified by the related effective slip. However, at large surfactant coverage, the effective shear plane crosses the electrostatic surface, and the presence of an effective stagnant layer explains that the zeta potential becomes lower than the ideal surface potential. 

Finally, we note that the difference between $\zsp$ and $\zes$ is on the order of 1\,nm or less. According to the discussion above, the zeta potential will significantly differ from the ideal surface potential when $|\zsp-\zes|$, the effective width of the stagnant layer or the effective slip length, is \new{not negligible compared to} the Debye length. Therefore, the effects reported here should be observed experimentally when using large salt concentration leading to nanometric Debye lengths, but at low salt concentration and large Debye length these molecular effects should become negligible and the ideal surface potential should provide an excellent approximation to the zeta potential. \new{To give an order of magnitude, one can estimate that the molecular effects observed here will become significant as soon as $|\zsp-\zes|/\debye \gtrsim 10\,\%$, corresponding to $\debye \lesssim 10$\,nm and $c_\text{s} \gtrsim 1$\,mM.}


\section{Conclusions}

In this work we have simulated the electro-osmotic flow of a film of aqueous electrolyte covered with a fluid surfactant layer and surrounded by a model gas. We measured the electro-osmotic velocity, defined as the difference between the bulk liquid velocity and the gas velocity, and computed the corresponding zeta potential as a function of surfactant coverage. We obtained a bell-shaped curve, with the zeta potential increasing with the surfactant coverage at first, and decreasing at large surfactant coverage. Furthermore, we compared the obtained zeta potential with the predictions of several models: first, we used the ideal surface potential as it is traditionally computed, accounting or not for ion binding; then, we used a more general model based only on Stokes equation, proposed by Huang et al. \cite{Huang2007,Huang2008}. In both cases, the models underestimated the zeta potential at low surfactant coverage, and overestimated it at large coverage. We then identified the weakness of Huang et al. model, which does not take into account the fact that the charges in the surfactants and bound ions are rigidly bound and do not transmit directly and locally the electrical force to the liquid. 
Considering the reciprocal streaming current configuration, we have derived an extended model accounting for the rigid body motion of the surfactants and bound ions. We have introduced an effective shear plane where the water+free ions and surfactants+bound ions velocity profiles meet in the SC configuration, different from the liquid-vapor slip plane.
\new{We have then proposed a new formula, Eq. \eqref{eq:zeta_new}, relating the zeta potential to the charge density profile of free ions and to the effective shear plane,} 
which can be used to predict the zeta potential of a liquid-vapor interface covered with fluid surfactants, once the equilibrium distribution of ions is known \new{(e.g. from short MD simulations)}. 
We have finally introduced the effective electrostatic surface in order to  give a simple interpretation for the discrepancy between the zeta potential of a liquid-vapor interface laden with a fluid surfactant layer, and the ideal surface potential, in terms of shift between the effective shear plane and the effective electrostatic surface. The shift between the two effective surfaces being on the order of the nanometer, the discrepancy between the zeta potential and the surface potential should only appear when the Debye length is also on the order of the nanometer, therefore at large salt concentration, and on the contrary the zeta potential should be described quite well by the ideal surface potential for large Debye lengths, i.e. at low salt concentration. \new{We discussed that deviations from existing models could become significant for $\debye \lesssim 10$\,nm, corresponding to $c_\text{s} \gtrsim 1$\,mM.} 

We hope this work will help experimentalists to better understand and control EK flows in liquid foams and soft nanofluidic systems involving surfactant-laden liquid-vapor interfaces. We should note also that our model predicts that the amplitude of the friction between the liquid and the vapor does not play a role. It would therefore be particularly interesting to test if this model can describe water-oil interfaces, which have also important industrial applications \cite{becher2001emulsions}, and which are also the subject of hot debates regarding the nature of their surface charge \cite{Marinova1996,Leunissen2007,Knecht2008,Vacha2011,Roke2012,Roger2012}. Similarly, our model could also in principle be used to describe the zeta potential of hydrophobic surfaces in the presence of large adsorbed molecules, e.g. surfactants, accounting for ion binding. Generally, this work also shows that electro-osmosis and streaming current, quantified by the zeta potential, result from the complex coupling between electrostatics and hydrodynamics at the interface, and cannot always be predicted by traditional models based on a static description of the interface. The crucial role of hydrodynamics, which has been highlighted also for other EK flows recently \cite{Ajdari2006,Lee2017}, could then be used to control and optimize EK flows in soft nanofluidic systems. 

\section*{Acknowledgments}

The authors thank D.M. Huang, L. Bocquet, C. Ybert, A.-L. Biance, C. Loison and M. Le Merrer for fruitful discussions. This work is supported by the ANR, projects ANR-13-JS09-0002-01 EFOAM and ANR-16-CE06-0004-01 NECtAR. 





\bibliography{library,library1,bibliorevue,bibliolaurent,bibliolaurent1,biblioadd,biblioadd1} 
\bibliographystyle{rsc} 

\end{document}